\documentclass[prb, aps, superscriptaddress, onecolumn,12pt]{revtex4}

\usepackage{bm}

\usepackage{amsmath}
\usepackage{amssymb}
\usepackage{amsfonts}
\usepackage{rotating}
\usepackage{wrapfig}
\usepackage{epic,color,curves,tikz}

\newcommand{\be}{\begin{eqnarray}}
\newcommand{\ee}{\end{eqnarray}}
\newcommand{\beno}{\begin{eqnarray*}}
\newcommand{\eeno}{\end{eqnarray*}}
\newcommand{\ba}{\begin{array}}
\newcommand{\ea}{\end{array}}
\newcommand{\la}{\langle}
\newcommand{\ra}{\rangle}
\newcommand{\no}{\nonumber}

\begin{document}

\pagenumbering{arabic}
\pagestyle{plain}

\author{Jianshu Cao}
\email{jianshu@mit.edu}
\affiliation{Department of Chemistry, Massachusetts Institute of Technology, Massachusetts, 02139 USA }

\author{Eli Pollak}
\affiliation{Chemical and Biological Physics Department, Weizmann Institute of Science,
76100 Rehovoth, Israel}

\title{Cavity-Induced Quantum Interference and Collective Interactions  \\ in van der Waals Systems }

\date{\today}

\begin{abstract}

The  central topic of this letter is to show that light-matter hybridization not only gives rise to novel dynamic responses but can also modify 
intermolecular interactions and induce new structural order.  Using
 the van der Waals (vdW) system in an optical cavity as an example,
we predict the effects of interference and collectivity in cavity-induced many-body dispersion interactions.  
Specifically,  the leading order correction due to cavity-induced quantum fluctuations leads to 3-body and 4-body 
 vdW interactions, which can align intermolecular vectors and are not pairwise additive. In addition, 
 the cavity-induced dipole leads to a single-molecule energy shift that aligns individual molecules, 
 and a pair-wise interaction that scales as  $R^{-3}$ instead of the standard $R^{-6}$ distance scaling.
 The coefficients of all these cavity-induced corrections depend on the cavity frequency 
 and are renormalized by the effective Rabi frequency, which in turn depends on the particle density. 
Finally, we study the interaction of the vdW system in a cavity with an external object 
and find a significant enhancement in the interaction range due to modified distance scaling laws. 
  These theoretical predictions suggest the possibility of cavity-induced nematic or smectic order  
 and may provide an essential clue to understand intriguing phenomena observed in optical cavities, 
 such as strongly-modified ground-state reactivity, ion transport
 and solvent polarity. 

\end{abstract}

\maketitle

\newpage

The van der Waals (vdW) interaction, also known as dispersion force,  is a fundamental force between any pair of atoms or molecules and 
plays an important role in physics and chemistry. \cite{novoselov16,sternbach21}  As illustrated in Fig.~1a, London demonstrated that 
quantum fluctuations in transition dipoles lead to a long-range pairwise attraction that scales with the distance as $R^{-6}$.\cite{london37,stone13}
The microscopic interpretation suggests that the vdW attraction can be manipulated by strong electromagnetic fields, which can be achieved
collectively in optical cavities.\cite{garcia21,thomas19,fukushima22,piejko23}  
Indeed, recent simulations of $H_2$ molecular systems\cite{haugland21,philbin22b}  
have demonstrated this possibility and suggested the exciting possibility of modifying the structure of molecular systems 
by tuning the light-matter coupling parameters.  These numerical studies of many-body systems are computationally expensive,
 whereas theoretical analysis has mostly been limited to pairwise interactions.\cite{milonni96,fiscelli20}
 In this paper, we aim to  reveal interference effects in cavity-induced many-body interactions and predict the scaling relations 
 of the resulting collective vdW interaction.    

This study is directly motivated by intriguing phenomena observed in optical cavities, including the significant modification of ground-state reactivity,
ion transport, and solvent polarity, which require better mechanistic understandings.\cite{thomas19,fukushima22,piejko23} 
Most theoretical studies have focused on polariton dynamics in unperturbed molecular configurations but cannot fully explain 
the cavity-catalyzed reactions\cite{galego19,li22,schafer21,cao203,mandal22,campos23,cao210}  
and other experiments.
As proposed in this and related studies, the collective vdW interactions in optical cavities can 
potentially modify the structure of vdW systems and change their thermodynamics and dynamic response.  Thus, our predictions 
can shed new light on how to  understand and control the unusual properties of condensed phase cavity systems.

{\it Model and basis sets} We consider an ensemble of two-level systems  or spins in an optical cavity.  Each spin represents a ground state and an excited state, which are coupled to the cavity mode.  Here, we consider the vdW system as electronic, but the same formalism applies to 
vibrations, rotations, and other degrees of freedom.  The molecular systems
interact via the dipole-dipole interactions (DDI) between fluctuating dipoles, which give rise to the vdW attraction.  The light-matter (LM) interaction in the cavity is described by the Pauli-Fierz (PF) Hamiltonian,\cite{power59} 
which contains the dipole self-energy (DSE) and counter-rotating-wave (CRW) terms. 
Thus, the overall Hamiltonian is given as
\be
H= H_{TCM} + H_{CRW}+  H_{DSE}+ H_{DDI}, 
\label{h}
\ee
where $H_{TCM}$ is the Hamiltonian of the Tavis-Cummings model (TCM),  
and other terms are defined later when they are evaluated.
Explicitly, the TCM Hamiltonian reads 
$
H_{TCM}=  E_c a^+ a +   E_m \sum_i   \sigma^+_i\sigma^-_i  +  H_{RWA}
$
where $a$ and $a^+$ are the lowering and raising operators of the cavity photon,   $\sigma_i^-$ and $\sigma_i^+$ 
are the lowering and raising operators
for the spin associated with the i-th molecule,   and $E_c$ and $E_m$ are their respective energies.  
The TCM  adopts  the rotating wave approximation (RWA) for the LM interaction, 
 $
	H_{RWA} = 	g ( \Sigma^+ a +  \Sigma^- a^+ ), 
$
where g is the LM coupling strength, $\Sigma^+ = \sum_i \sigma_i^+$ and $\Sigma^- = \sum_i \sigma_i^-$.

The vdW attraction arises from  the energy shift on the ground state due to the DDI and involves zero or double quantum transitions.
Therefore, it suffices to consider the eigenstates of the TCM on the ground and  double-excitation manifolds and evaluate other terms 
in Eq.~(\ref{h})  via perturbation.   
Following the detailed derivation given in the Supporting Information (SI), 
we first construct the collective particle states and then diagonalize  $H_{RWA}$ in each excitation manifolds to obtain LM hybrid states.  
Some aspects of the basis set construction and its application to cavity polaritons 
can be found in literature.\cite{dicke54,gross82,wersall19,delpo20,cederbaum22,moiseyev22}

{\it Dipole-dipole interaction (DDI)}  To begin,  we introduce the dipole-dipole interaction 
$
H_{DDI} = - \sum_{i>j} \sigma_i T_{ij} \sigma_j, 
$
where  $\sigma = \sigma^+ + \sigma^-$ 
and $T$ is the projection of the DDI tensor in the polarization direction of the 
cavity field. 
Since two particle operators are involved, $H_{DDI}$ connects 
the ground state and the double-excitation manifold.   For simplicity,  we use $M_2$ to denote the double-excitation manifold,
i.e., $M_2$; $\lambda$ to specify an eigen-state in $M_2$; and $E_{2,\lambda}$
to specify the corresponding eigen-energy.  Then, the second-order perturbation calculation of the ground state energy gives
\be
E_{DDI} =	-\sum_{\lambda} { {|\left<M_{2,\lambda} |H_{DDI} |G \right>|^2} \over E_{2,\lambda} }  
	=	-\sum_{\lambda} {{|\left<M_{2,\lambda} |H_{DDI} | G\right>|^2} \over { 2\omega } }
	-	\sum_{P_2} \left( {1\over E_{2,P} } - {1\over 2\omega }\right)
	 \left|\left<P_2 | H_{DDI} | G \right>\right|^2 \no
\label{dipole}
\ee
where $P_2$ represents the bright polariton states in $M_2$ that have energies $E_{2,P}$ which differ from the uncoupled molecular energy, $2\omega$. 
By virtue of the completeness relationship, $\sum_{\lambda} |M_{2,\lambda} \ra \la M_{2,\lambda} |= \hat{I}$,   
the first term in Eq.~(\ref{dipole}) reduces to
\be
	E_{VDW}= -\sum_{\lambda} {{|\left<M_{2,\lambda} |H_{DDI} |G\right>|^2} \over {2\omega} }
	= - {1\over{2\omega}} \sum_{i>j} T_{ij}^2
\label{vdw}
\ee
which recovers the standard vdW attraction and is isotropic (see Fig.~1a).   
The cavity-induced effect arises from the second term in Eq.~(\ref{dipole}), i.e., the contribution of the (bright) polariton manifold, $P_2$, 
which consists of two sets of states,  $P_{2,\pm}$ and $P^{\mu}_{2,\pm}$ (see the SI). 
First, the contribution to the energy shift coming from the polariton states $P_{2, \pm}$ is
\be
-\Delta E^{(1)}_P &=&  \sum_{\pm} \left( {1\over E_{2,\pm} } - {1\over 2\omega }\right)
	 \left|\left< P_{2,\pm}  | H_{DDI}  | G \right>\right|^2   
=	{\Omega^2_{4N-2} \over \omega (4\omega^2-\Omega^2_{4N-2}) }
{1 \over {N(2N-1)}}    |\sum_{i>j} T_{ij} |^2  
\label{4body} 
\ee
where $\Omega_{4N-2}=\sqrt{4N-2}g$ is the effective Rabi frequency. 
Next, the contribution from $P^{\mu}_{2,\pm}$ is
\be
-\Delta E^{(2)}_{P} =  \sum_{\mu,\pm} \left( {1\over E^{\mu}_{2,\pm} } - {1\over 2\omega }\right)
	 \left|\left<P^{\mu}_{2,\pm}  | H_{DDI} | G \right>\right|^2  
= {\Omega^2_{N-2} \over \omega (4\omega^2-\Omega^2_{N-2}) }
{1 \over {2(N-2)}}    \sum_{i,j,k} T_{ij} T_{jk}
\label{3body}
\ee
where $\Omega_{N-2}=\sqrt{N-2}g$ is the effective Rabi frequency.

As illustrated in Fig.~1d and Fig.~1c, Eqs.~(\ref{4body})  and (\ref{3body}) represent the cavity-induced 4-body and 3-body vdW interactions, respectively,
 and demonstrate the quantum nature of light-matter hybridization:
 (i)   Without collective coupling to the cavity, these many-body terms cannot appear in the second-order perturbation, 
 but require higher order treatment.\cite{cao6}
 (ii) The pre-factors of these cavity-indued terms take the form 
 \be
\mbox{prefactor} = { \Omega_N^2 \over \omega^2 -c \Omega_N^2}
\label{prefactor} 
\ee
 which is renormalized by the effective Rabi frequency and is thus non-addictive in the molecular density. 
  (iii) Since the predicted effect depends on the ratio $\Omega_N / \omega$,  
 lower frequencies under vibrational strong coupling (VSC) can lead to larger enhancement than 
the coupling to electronic transitions.   
(iv)  As discussed in the SI, the magnitudes of these contributions depend on the structure of the sample and scale differently with the molecular 
density in the gas, liquid, and solid phases. 

  \begin{figure}
\centerline{\scalebox{0.6}{\includegraphics[trim=0 0 0 0,clip]{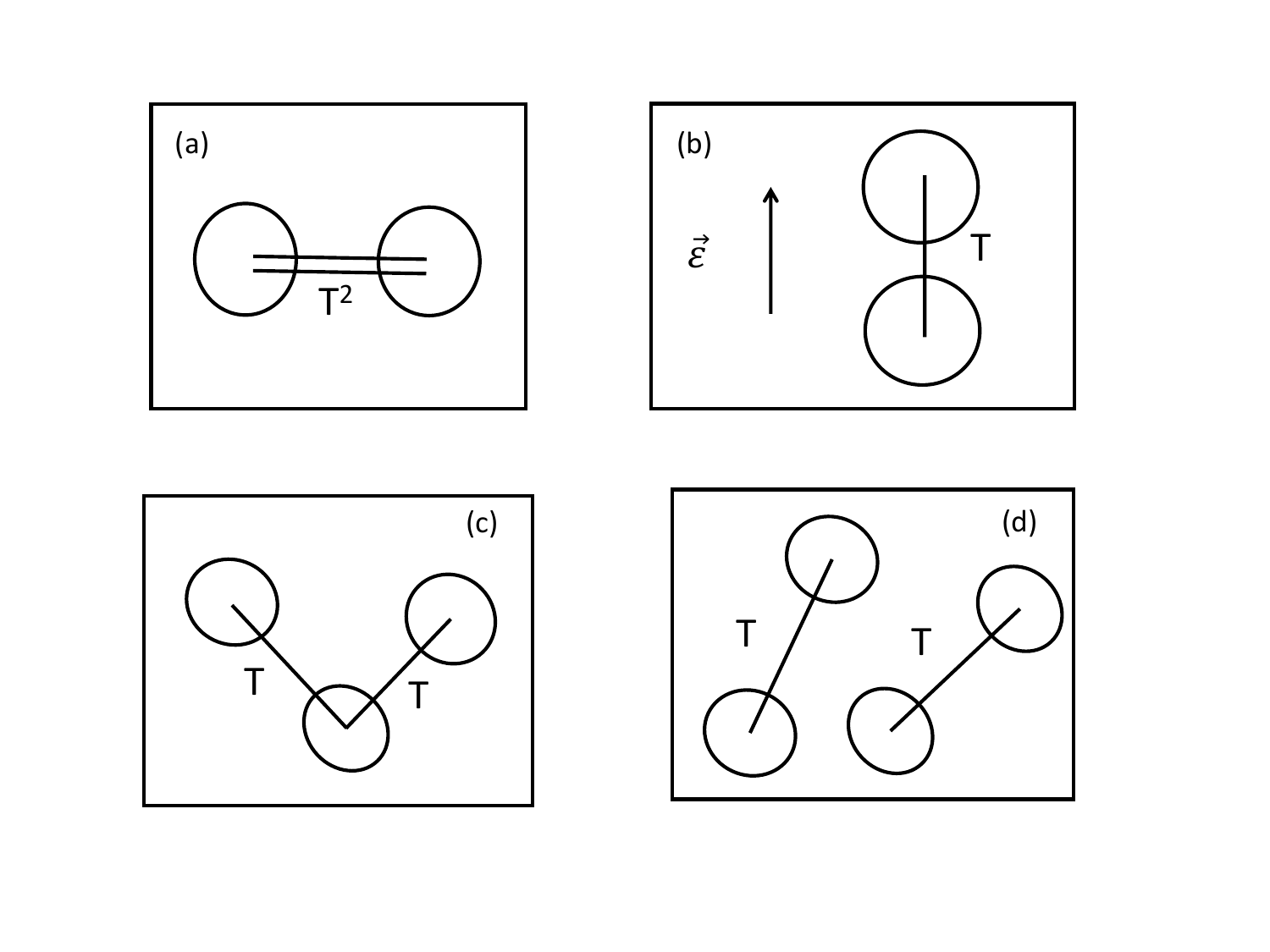}}}
\caption{Illustration of the various contributions to the vdW interaction in an optical cavity:
(a) the standard isotropic and pairwise vdW interaction;
(b) the pairwise  interaction between cavity-induced dipoles with a distance scaling of ${ 1/ R^3}$;
(c) the 3-body contribution;  (d) the 4-body contribution.  Both the 3-body and 4-body terms arise from
the cavity-induced quantum interference in dipole fluctuations.  }
\end{figure}

{\it Counter-rotating-wave term (CRW) } 
The TCM is the RWA to the Dicke model\cite{dicke54,spano90}, and 
the difference between the two models is the CRW term, 
$
H_{CRW}  = g (a^+ \Sigma^+ + a \Sigma^-). \no
$
The first contribution is 
\be
	-E^{(1)}_{CRW}	 =	\sum_{\pm}
			\frac{\left|\la P_{2,\pm} | H_{CRW} |G\ra\right|^2}{E_{2,\pm}} 
			=   \frac{2\omega \Omega^2_{N}} {(2\omega)^2 - \Omega^2_{4N-2}} 
\label{crw1}
\ee
which introduces a one-body energy shift and can potentially 
lead to an orientational distribution with respect to  the polarization direction of the cavity field.  
The second contribution arises from the cross term between 
  $H_{CRW}$ and $H_{DDI}$ and is given explicitly as
\be
	-E^{(2)}_{CRW} = 2 \sum_{\pm} \frac{\la G| H_{CRW} |P_\pm\ra\la P_\pm| H_{DDI} | G \ra}{ E_{2,\pm}} 
		=	\frac{4  g^2}{(2\omega)^2 - \Omega^2_{4N-2}} \sum_{i>j} T_{ij} 
\label{crw2}
\ee
This term is pairwise with a non-additive pre-factor and a distance scaling of ${ 1/ R^3}$,  which 
results from the interaction of two dipoles induced by the cavity field (see Fig.~1b).

{\it Dipole self-energy (DSE)}
The  PF Hamiltonian also contains the dipole self-energy term in the form of
$
H_{DSE} =  \Sigma^2  {g^2 / \omega_c},
$
where $\Sigma =    \sum_i (\sigma^-_i + \sigma^+_i )$ is the collective transition dipole. 
First-order perturbation evaluation of $H_{DSE}$  yields the same form as in $E^{(1)}$ in Eq.~({\ref{crw1})
but with opposite sign.  As a result, the two single-particle energy shifts  due to $H_{DSE}$ and $H_{CRW}$ 
cancel exactly in the far off-resonance regime of $\omega_c \ll \omega_m $ and cancel by half at resonance $\omega_c = \omega_m $.  
The second contribution arises from the cross term between  $H_{DSE}$ and $H_{DDI}$ and 
 is identical to Eq.~(\ref{crw2}), i.e., the second contribution of the CRW Hamiltonian.  
A detailed derivation can be found in the SI. 
 
{\it Cavity frequency dependence}
The standard vdW attraction is a non-resonant effect, which is independent of the cavity frequency, as evidented in Eq.~(\ref{vdw}). 
In comparison, the cavity-induced corrections depend on the cavity frequency, introducing an additional control parameter.  
The above calculations are limited to the resonance case of $\omega_m = \omega_c =\omega$, which will now be generalized  
to the off-resonance case of $\omega_m  \neq \omega_c$.   For simplicity, we focus on the  $P_2^{\mu}$ contribution to the energy shift, $\Delta E_P^{(2)}$,
which dominates the the many-body interference effect.  Then,  Eq.~(\ref{3body}) is generalized to 
\beno
-\Delta E^{(2)}_{P} =
 \sum_{\mu,\pm} \left( {1\over E^{\mu}_{2,\pm} } - {1\over 2\omega_m }\right)
	 \left|\left<P^{\mu}_{2,\pm}  | H_{DDI} | G \right>\right|^2   
=  [ { 2\omega_m \over (2\omega_m+\delta)^2 -\Delta_{N-2}^2 } - { 1 \over 2 \omega_m} ] \sum_{\mu} |\la s^{\mu}_2 | H_{DDI} | g \ra|^2 
 \eeno
 where the frequency detuning is $\delta= { (\omega_c - \omega_m ) / 2} $, 
  $\Omega_{N-2}=\sqrt{N-2} g$,  and $\Delta_{N-2} = \sqrt{ \Omega_{N-2}^2 + \delta^2} $.
In comparison with the resonant result in Eq.~(\ref{3body}), positive detuning reduces the pre-factor and 
thus suppresses the interference effect, whereas  negative detuning does the opposite and enhances the interference effect.  
As we further increase the cavity frequency such that $\delta >\delta_c$, with the crossover detuning
$
\delta_c = { \Omega^2_{N-2} \over 4 \omega_m} ,
$
then $\Delta E^{(2)}_P \ge 0$. As a result, 
 the cavity-induced  non-additive interaction can reduce the vdW attraction or even change attraction to repulsion.    
 This prediction is consistent with the dramatic effects of frequency detuning reported in a recent {\it ab initio} simulation.\cite{haugland21} 

  \begin{figure}
\centerline{\scalebox{0.6}{\includegraphics[trim=0 0 0 0,clip]{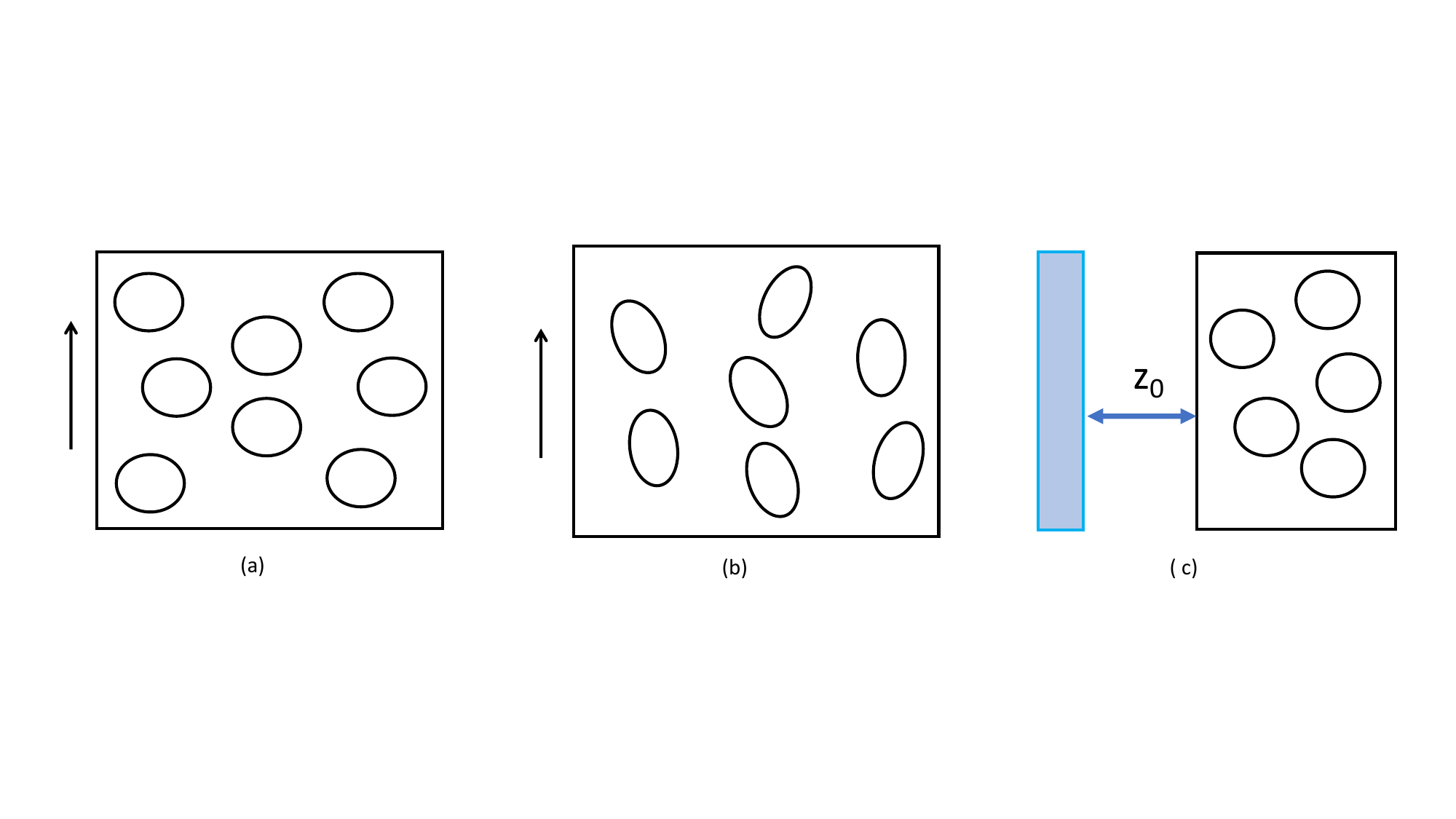}}}
\caption{Illustration of various structural orders in condensed phase vdW systems:
(a)  intermolecular orientational order of spherical particles;
(b) orientational order of non-spherical particles; 
(c)  collective attraction between a cavity vdW system and an external slab.}
\end{figure}

{\it Orientational order}
The scalar formulation presented thus far should be understood as a projection into the polarization of the cavity field 
and can predict structural changes in cavity systems.  Specifically, 
 the LM interaction and resulting hybrid states are all defined in the polarization direction of the cavity field (i.e., the z axis):
 The transition dipole is the projection along the polarization, $\mu_z$,  the dipole-dipole interaction is the z-projection of the dipole tensor 
 ${\cal T}$, 
 $
T=  \mu_z {\cal T}_{zz}   \mu_z \no,
$
and  the coupling strength to the cavity field is also defined similarly as
 $
 g  =   \mu_z   g_0 /\mu  \no
 $
 where $g_0$ is the maximal coupling constant, and $\mu = |\vec{\mu}| $ is the magnitude of the transition dipole.
 In the perpendicular directions, the molecular states are unperturbed. As a result, the leading order
 in the vdW attraction in Eq.~(\ref{vdw}) recovers the  isotropic vdW potential, whereas
 the cavity-induced modifications  are along the 
 polarization direction of the cavity field and thus break the isotropic distribution of the cavity-free molecular sample.  
 Consider the following different cases:

 \begin{enumerate}
\item
 Spherical particles have no orientational preferences, so the orientational order exists not on individual particles but
 between particles. The cavity-induced many-body polarization effects
  align the intermolecular axis and can potentially lead to an orientational order, similar to the smectic phase in liquid crystals (see Fig.~2a). 
 \item
  An interesting scenario arises when the spherical symmetry of individual particles is broken, as in most molecules. 
  The principal molecular axis indicates the preferential polarizability direction, and the transition dipole is defined in 
 the molecular frame as $\vec{\mu}_i (\hat{\Omega}_i)$,  
  where $ \hat{\Omega}_i $ is the solid angle associated with the i-th molecule.\cite{cao8} 
As a result, the cavity field can align both the molecular axis on the single-particle level 
and the intermolecular axis on the ensemble level, creating both the nematic and 
smectic orders (see Fig.~2b).\cite{fiedler18,stemo22,piejko23}
\item
In addition to induced dipoles, most molecules possess permanent dipoles, which will reinforce both 
the dipole-dipole interaction and the light-matter interaction.  The permanent dipoles in 
polarizable systems have been studied within the Drude oscillator model and show a prominent effect in
the dielectric constant and dielectric response.\cite{cao9}  The light-matter interaction in polarizable dipolar systems
is an interesting topic for future study. 
\end{enumerate}

{\it vdW interaction with external objects}
In addition to the homogenous polarizable system considered so far,  another interesting scenario is the vdW 
interaction with an external object, such as a mirror in the Fabry-Perot (FP) cavity or a surface in a plasmonic cavity. 
The presence of such an external object breaks the spatial homogeneity and naturally reinforces the interference effect in the vdW 
interaction.  
As an illustrative example, we consider the coupling between a 2D molecular thin film and a molecular sample in the cavity separated 
by a distance $z_0$, as illustrated in Fig.~2c.   
Assuming that the spatial extension of the thin film is much larger than $z_0$, 
we integrate over the area of the thin film and obtain the scaling relationships,
\beno
-E_{vdW} \propto \sum_i |T_{0i}|^2  &  = & \int\int dx  dy { \rho_2 \over ({x^2 + y^2 +z_0^2})^3 }  \propto {\rho_2 \over z_0^4}  \\
-E^{cav}_{vdW} \propto  |\sum_i T_{0i} |^2 &  = & \left[ \int dx \int dy { \rho_2 \over (x^2 + y^2 +z_0^2)^{3/2} }  \right]^2 \propto {\rho_2^2 \over z_0^2} 
\eeno
where $\rho_2$ is the molecular density of the 2D plane.  As suggested by the different scalings with respect to $z_0$,
   the cavity-induced vdW interaction has a  much longer interaction range than the standard vdW attraction.  
This effect is related to the cavity Casimir-Polder forces and generally depends on the shape and dimension 
of the external object.\cite{thiyam15,galego19,karimpour22}
  
Further, we comment on the potential role of  the enhanced attraction to the cavity surface in cavity-catalyzed reactions. 
With the enhanced range and strength, the cavity wall or mirror can attract reactive molecules in an optical cavity and thus create 
a high density molecular layer. The surface-bound molecule can potentially modify its structure and thus change its reactivity. 
As suggested in Eq.~(\ref{prefactor}), the cavity effect scales generally as $\Omega^2/\omega^2$ and is thus stronger  
for the vibrational coupling than for the electronic coupling, so the surface effect can be relevant in the vibrational 
strong-coupling (VSC) regime.\cite{galego19,li22,cao210}  

 {\it Conclusion} 
In summary,  with a second-order perturbation on the hybrid light-mater state basis, we predicted the coorperativity and interference of 
 vdW interactions in optical cavities:  (i)  In addition to the standard pairwise interaction, the dipole fluctuations in cavities lead to 3-body and 4-body 
 vdW interactions, which increase with molecular density and  can align intermolecular vectors. (ii) Both the DSE and CRW terms
 contribute  constant energy shifts on the single-particle level  and pairwise interactions between cavity-induced dipoles. 
 Depending on the relative orientation, the latter interaction can be repulsive or attractive and scales with the distance as $R^{-3}$. 
 (iii) The pre-factors of these cavity-induced interactions depend on the cavity frequency and scale with molecular density and light-matter coupling
 nonlinearly. (iv)  For the interaction with an external object, such as a wall or mirror, the interference effect leads to distinct distance scaling laws
 and significantly enhances the interaction range and strength. 
 These predictions suggest the possibility of structural changes, in particular, the emergence of nematic or smectic order, and may help explain 
 recent cavity experiments. 
 
 We now commend briefly on the relevance of the cavity-induced effects in realistic systems: 
 (i) The leading-order  coefficient of the many-body terms in Eqs.~(\ref{3body}) and (\ref{4body})  scales with the ratio of 
 $\Omega_N^2/\omega_m^2$, which can reach the range of 0.1 in the strong coupling regime.
  (ii) The collectivity in the LM interaction leads to non-linear N-scaling in Eq.~(\ref{prefactor}), 
  which has been observed in simulations.\cite{haugland21,philbin22b}
  (iii) The $R^{-3}$ term in Eqs.~(\ref{crw2}) has been analyzed 
  on the pairwise level \cite{fiscelli20,philbin22b} and can be further 
  enhanced by the  collective correction.     
 
 Though the current analysis is devoted to the vdW interaction in optical cavities, 
the basic idea that optical cavities can modulate intermolecular potentials  collectively is general and can be applied 
to other types of interactions including hydrogen bonding.  
As a result, the unusual dynamical properties in cavities can arise not only from the dynamical response due to the light-matter
hybridization, but also from the structural modifications due to the cavity-induced intermolecular interactions.  
Further theoretical study can extend the current analysis from a single-photon mode 
to a multi-mode cavity,\cite{cao211} which leads to
 the longitudinal component and retardation in the dipolar coupling; however,
 the interference and collective effects predicted here are expected to remain effective.   
Another  direction to explore is the cavity-modified  interactions
under VSC, which may help understand the intriguing discovery 
of cavity-catalyzed chemical reactions.

\section*{Acknowledgement}

This work was generously supported by the NSF (Grants No. CHE 1800301 and No. CHE1836913) and the MIT Sloan Fund. 
E. Pollak acknowledges  a joint Israel Science Foundation, Natural National Science Foundation of China grant 2965/19.
J. Cao acknowledges the sponsorship of the Rosi and Max Varon Visiting Professorship at the Weizmann Institute of Science and 
 the Marie Curie FRIAS COFUND Fellowship Programme (FCFP) during his sabbatical in Germany.
 
 \newpage
 
\bibliography{polariton,pub_cao}

\end{document}